\documentclass[twocolumn,superscriptaddress,showpacs,showkeys,preprintnumbers,amsmath,amssymb]{revtex4}

\usepackage{graphicx}
\usepackage{dcolumn}
\usepackage{bm}
\usepackage{longtable}
\begin{document}

\title{Transport Coherence in Frictional Ratchets}
\author{Raishma Krishnan}
\affiliation{Institute of Physics, Bhubaneswar-751005, India}
\email{raishma@iopb.res.in, jayan@iopb.res.in}
\author{Debasis Dan}
\affiliation{Department of Physics, Indiana University, Bloomington 
47405, U.S.A}
\email{ddan@indiana.edu}
\author {A. M. Jayannavar}
\affiliation{Institute of Physics, Bhubaneswar-751005, India}

\begin{abstract}
Abstract: We study the phenomena of noise induced transport in frictional 
ratchet systems. For this we consider a Brownian particle moving in 
a space dependent frictional medium in the presence of external white noise 
fluctuations. To get the directed transport, unlike in other ratchet models 
like flashing or rocking ratchets etc., we do not require the 
potential experienced by the particle to be asymmetric nor do 
we require the external fluctuations to be correlated. We have 
obtained analytical expressions for current and the diffusion coefficient. 
We show that the frictional ratchets do not exhibit a pronounced 
coherence in the transport in that the diffusion spread 
overshadows the accompanying directed transport in system with 
finite spatial extensions.
\end{abstract}


 \maketitle
 \section{Introduction}
 Ratchets, also termed as Brownian motors, are systems 
 that exploit the nonequilibrium 
 fluctuations that are present in the medium to generate directed 
 flow of Brownian particles in the absence of any net 
 external force or bias \cite{reiman, 1amj}. Recently, 
 such Brownian motors have been studied extensively because they 
 are believed to share common features with biological motors. 
 Biological protien motors convert the chemical energy into 
 macroscopic work i.e., they transport cargo efficiently with 
high reliablility at room temperature in the presence 
of a very noisy environment 
 \cite{ptoday,resonance}. 

  In thermal equilibrium, Onsager's principle of detailed balance,  
 where the forward transition in any pathway 
 is on average balanced by an identical transition in the backward direction, 
holds due to which there cannot be any particle current in the system. Also, 
 an understanding of the second law of thermodynamics \cite{feyn} 
 tells us that a system when left to itself in thermal equilibrium would not 
 cause any preferential motion even in the presence of spatial 
 anisotropy. Thus external nonequilibrium fluctuations are 
 ubiquitous to drive a net directed flow \cite{reiman,1amj}. 

 Much has been studied from the viewpoint of statistical physics in 
 ratchet models to understand how unidirectional motion emerges 
 from nonequilibrium fluctuations \cite{reiman}. Several physical models like 
 flashing ratchets, rocking ratchets, time asymmetric ratchets etc.,
 where potential has been taken to be asymmetric in space, 
 have been developed. In these models to generate noise induced 
 directional transport the nonequilibrium fluctuations have to be 
 time correlated. In an earlier work it was shown that 
 in the case of inhomogeneous systems, where the friction coefficient and/or 
 temperature varies in space \cite{buttiker}, it is possible to 
 get unidirectional currents even in a symmetric potential in 
 the absence of a net bias. Moreover, external fluctuations 
 need not be correlated in time  \cite{amj,pareek}. 
 Such ratchets are termed as 
 frictional ratchets. In the presence of external parametric noise 
 the particle on an average absorbs energy from the noise source. 
 The particle spends larger time in the region of space where the 
 friction is higher and hence the energy absorption from the noise 
 source is higher in these regions. Therefore, the particle in the higher 
 friction regions feel effectively higher temperatures. Thus, the problem 
 of particle motion in an inhomogeneous medium in presence of 
 an external noise becomes equivalent to the problem in a space 
 dependent temperature. Such systems are known to generate 
 unidirectional currents \cite{1amj,pareek,buttiker,land}. 
 This  follows as a corollary to 
 Landauer's blow torch theorem that the notion of stability changes 
 dramatically in the presence of temperature inhomogenieties. 
 In such cases the notion of local stability, valid in equilibrium 
 systems, does not hold. Frictional inhomogenieties are 
 common in superlattice structures, 
 semiconductors or motion in porous media. It is believed that 
 molecular motor protiens moving close along the periodic structures 
 of microtubules  experience a space dependent friction. Frictional 
 inhomogenity changes the dynamics of the particle nontrivially as 
 compared to the homogeneous case. This in turn has been shown to 
 give rise to many counter intuitive phenomena in driven 
non-equilibrium systems \cite{thesis}.

Considerable amount of work has been done on the nature of current, 
current reversals as well as thermodynamic efficiency in these systems. 
 However, the question of reliability or the coherence of transport has 
 recieved only little attention \cite{low,sch,land}. Transport of 
 Brownian particles are always accompanied by a diffusive spread. 
 The coherence of transport means a large particle current 
 accompanied by a minimal diffusive spread. When 
a particle on an average moves a distance $L$ due to 
its velocity, there will always be an accompanying 
 diffusive spread. If this diffusive spread is much smaller than the 
 distance travelled, then the particle motion is considered to be coherent 
 or optimal or reliable. This is in turn quantified by a 
 dimensionless quantity, Pe$\acute{c}$let 
 number Pe, which is the ratio of current to the diffusion constant. Higher 
 the Pe$\acute{c}$let number, more coherent is the transport.
 The Pe$\acute{c}$let numbers for some of the models studied 
 show low coherence of transport ($Pe=0.2$) \cite{low}. 
 But experimental studies 
 on molecular motors showed more reliable transport with Pe$\acute{c}$let 
 number ranging from 2 to 6 \cite{high}. Thus study of transport coherence is 
 of importance in identifying a proper model for the 
biological motors and also in nanoscale particle separation 
devices based on current reversals.

 In this work we study the coherence of transport of a Brownian particle in a
 medium with frictional inhomogeneity and an external 
 parametric white noise. We show that this system does not 
 exhibit pronounced coherence. 
 \section{The Model:}
 We consider the overdamped dynamics of a Brownian particle moving 
 in a medium with spatially varying frictional coefficient $\eta(q)$ at 
 temperature $T$. Using a  microscopic treatment the 
 Langevin equations for the Brownian 
 particle in a space dependent frictional medium has been obtained 
 earlier \cite{amj,pareek}. The corresponding overdamped Langevin 
 equation of motion is given by
 \begin{equation}
 {\dot{q}} = {- {\frac{V^\prime}{\eta(q)}}} - {\frac{k_BT \eta^{\prime}(q)}{2
 {[\eta(q)]}^2}} + {\sqrt {\frac {k_BT}{\eta(q)}}}f(t)
 \end{equation}
 with $<f(t)> = 0$, and $ <f(t)f(t^\prime)> = 2 \delta(t-t^\prime)$ 
 where $<...>$ 
 denotes the ensemble average and $q$ the coordinate of the particle.

 The system is then subjected to an external parametric additive 
 white noise fluctuating force $\xi(t)$, so that the equation of motion becomes
 \begin{equation}
 \dot{q} = {- \frac{V^\prime}{\eta(q)}} - \frac {k_BT {\eta^\prime}
 (q)}{ 2{[\eta(q)]}^2} + {\sqrt{\frac{k_BT}{\eta(q)}}f(t)} + \xi(t)
 \end{equation}
 with $<\xi(t)>=0$ and $<\xi(t) \xi(t^\prime)> = 2 \Gamma \delta(t-t^\prime)$,
 where $\Gamma$ is the strength of the external white noise $\xi(t)$. 
 The corresponding Fokker-Planck equation is given by \cite{riskin}
 \begin{equation}
 \frac {\partial P}{\partial t}= \frac {\partial}{\partial q} 
 \left[{\left\{\frac{V^\prime(q)}{\eta(q)}\right\}P+ \left\{\frac{k_BT}
 {\eta(q)}+\Gamma\right\}\frac{\partial P}{\partial q}}\right]
 \end{equation}
 For periodic functions $V(q)$ and $\eta(q)$ with periodicity $2\pi$, 
 one can readily obtain analytical expression for current 
 and is given by \cite{amj,pareek}
 \begin{equation}
 J= \frac{1-exp\,[{-2\,\pi\,\delta}]}{\int_0^{2\pi} dy exp\,[-\psi(y)] 
 \int_{y}^{y+2\pi}dx \frac{exp\,\,[{\psi(x)}]}{A(x)}}
 \end{equation}
 with the generalized potential $\psi(q)$ as
 \begin{equation}
 \psi(q)=\int^{q} dx \frac{V^\prime(x)}{k_BT+\Gamma \eta(x)}
 \end{equation}
 and  $A(q)$ as
 \begin{equation}
 A(q)=\frac{k_BT+\Gamma\eta(q)}{\eta(q)}\\
 \end{equation}
 with
 \begin{equation}
 \delta = \psi(q)-\psi(q+2\pi)\nonumber
 \end{equation}
 which inturn determines the effective slope of the generalized 
 potential $\psi(q)$. Hence the sign of $\delta$ gives the 
 direction of current which follows from Eqn(4). 

 In our present work we have taken the potential
$V(q)=V_0[1-cos(q)]$ and $\eta(q)=\eta_0[1-\alpha cos(q-\phi)]^{-1}$, 
$0\,<\,\alpha\,<\, 1$. 
 For simplicity we have restricted to the case where $T=0$. For this case 
 the effective potential $\psi(q)$, $\delta$ and $A(q)$ are given by
 \begin{eqnarray}
 \psi(q)&=& \frac{V_0}{\Gamma \eta_0}[1-cos(q)+ \frac{\alpha}
 {4}[cos(2q-\phi)-cos(\phi)]\nonumber\\
 &-&  
 \frac{\alpha}{2} q sin(\phi)]\\
 \delta &=& \frac{V_0 \pi \alpha sin(\phi)}{\Gamma \eta_0}\\
 A(q) &=& \Gamma
 \end{eqnarray}
 Following references \cite{rec}, one can obtain exact analytical expressions 
 for the diffusion coefficient $D$ and the average current $J$ as
 \begin{equation}
 D=\frac{\int_{q_0}^{q_0+L}\frac{ dx}{L}\,A(x)\, {[I_+(x)]}^2 I_-(x)}{\left[{\int_{q_0}
 ^{q_0+L}\frac{dx}{L}I_+(x)}\right]^3}
 \end{equation}
 \begin{equation}
 J= L\frac{1-exp\,[-L \delta]}
 {{\int_{q_0}^{q_0+L}\frac{dx}{L}I_+(x)}}
 \end{equation}
 where $I_+(x)$ and $I_-(x)$ are as given below
 \begin{eqnarray}
 I_+(x)&=& \frac{1}{A(x)}\,\,exp\,[\psi(x)]\,\int_{x-L}^{x} dy 
 \,\,exp\,[-\,\psi(y)]\\
 I_-(x)&=& exp\,[- \psi(x)] \int_{x}^{x+L} dy\,\, 
 \frac{1}{A(y)}\,exp\,[\psi(y)]
 \end{eqnarray}
 $L$ here represents the period of the potential ($=2 \pi$ in our case).
 Now, the time taken for a Brownian particle to travel a distance 
 $L$ is given as $\tau=L/v$ and the spread of the particle in the same time is 
  given as $<(\Delta q)^2>=2D\tau$. For a reliable transport we require 
 $<(\Delta q)^2>=2D\tau < L^2$. This in turn implies that $Pe=Lv/D >2$ 
 for coherent transport. 

 \section{Results and Discussions}
 We study the current ($J$), diffusion constant ($D$) and the 
 Pe$\acute{c}$let number ($Pe$) as a function of the phase difference $\phi$ 
 between the periodic functions $V(q)$ and $\eta(q)$, noise 
 strength, $\Gamma$ and $\alpha$, 
 the amplitude of the periodic modulation of $\eta(q)$. 
 All the physical quantities are taken in dimensionless form. 

 In Fig. (1) we plotted $J$, $D$ and $Pe$ as a function of 
 phase $\phi$ by numerically integrating Eqn 9 $\,\&\,$ 10 
 using quadrature methods. All other physical parameters are 
 mentioned in the figure caption. As expected, 
 all the quantities are periodic functions of phase $\phi$ with the 
 current $J$ being zero at $\phi=0\,\&\,2\pi$ \cite{pareek,d1}. The 
 Pe$\acute{c}$let number $Pe$ exhibits a maximum with 
 the maximum value being $1$. Thus in the parameters 
 considered in Fig. 1 the transport is not coherent.   

 In figures 2 and 3 we have plotted $J$, $D$ and $Pe$ as a 
 function of $\alpha$ and $\Gamma$ respectively for some typical values of 
 physical parameters given in the figure captions. It is clear from 
 these figures that in the parameter range we have considered the 
 transport is not very coherent. However, the transport coherence 
 in our model is found to be much larger than those in the 
 earlier nonfrictional ratchet models  where the Pe$\acute{c}$let number 
 is always less than $0.2$ \cite{low}. 
\begin{widetext}
 \begin{figure}[hbp!]
 \begin{center}
\input{epsf}
\hskip15cm \epsfxsize=4.1in \epsfbox{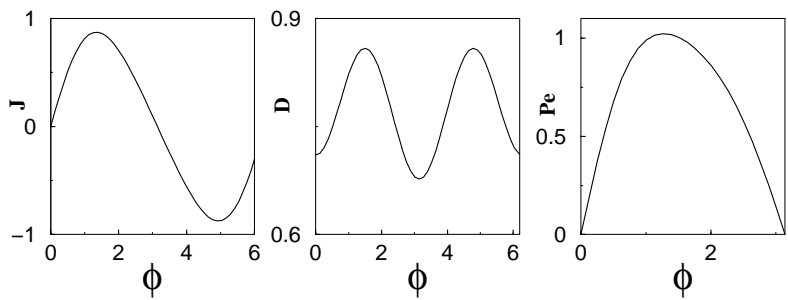}
 \caption{Plot J, D and Pe for $\Gamma=1\,\,, \, \alpha=0.5$}
 \end{center}
 \end{figure}
\end{widetext}

\newpage
 \begin{widetext}
\begin{figure}[htp!]
 \begin{center}
 \input{epsf}
 \hskip15cm\epsfxsize=4.8in \epsfbox{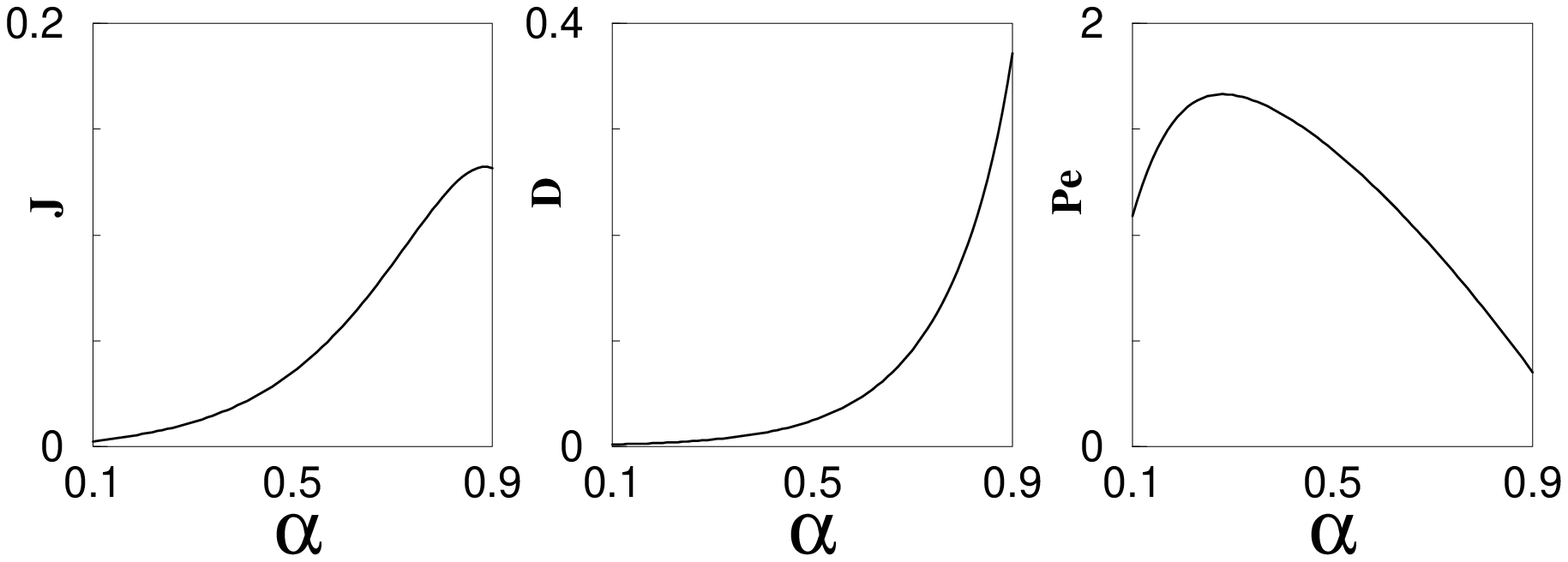}
 \caption{Plot of J, D, and Pe for $\phi=\pi/2 \,,\, \Gamma=1$}
 \input{epsf}
 \hskip15cm \epsfxsize=4.8in \epsfbox{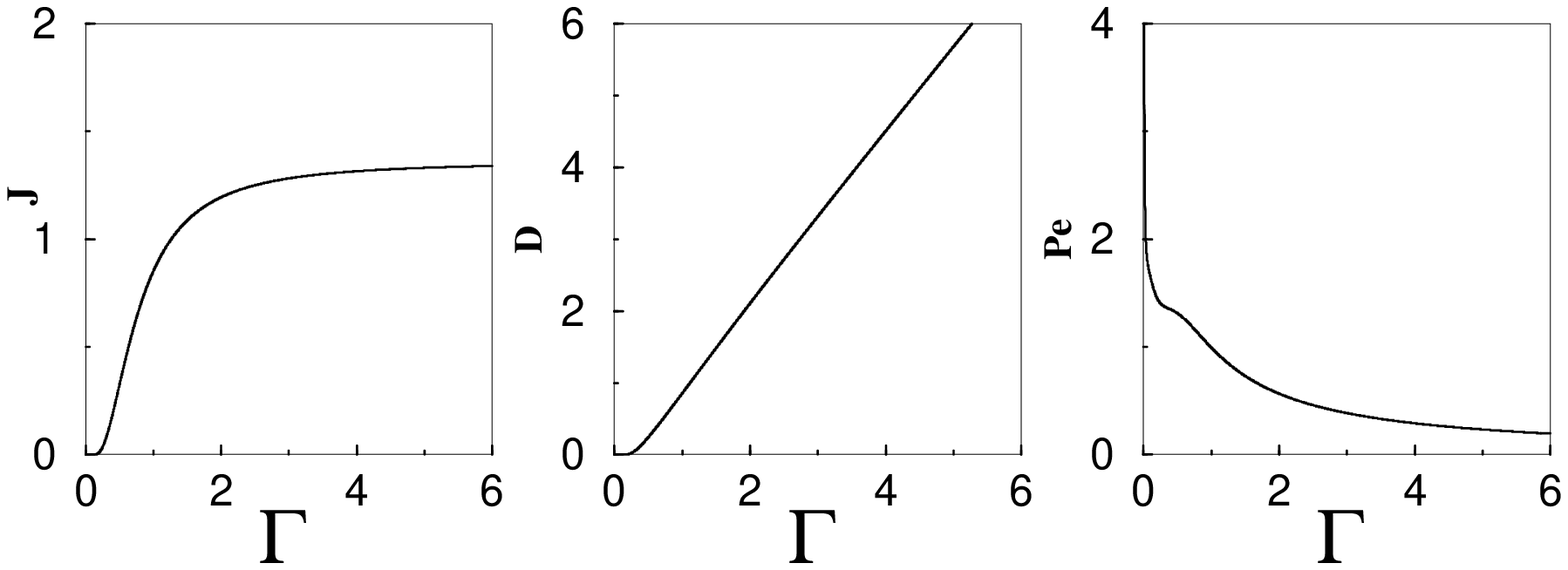}
 \caption{Plot of J, D and Pe for  $\phi=\pi/2 \,,\, \alpha=0.5$}
 \end{center}
 \end{figure}
 \end{widetext}
 \section{Conclusions}

 We have studied the reliability of transport in frictional ratchet systems 
 in the presence of external white noise 
 fluctuations. We restrict our study to the simple case where the 
 temperature of the heat bath is zero (absence of thermal noise). 
 The transport coefficients, both the current as well as diffusion,  
 arises solely due to the presence of external noise. 
 Our analysis indicates that for this special case, the 
 transport is not very coherent. However, the obtained $Pe$ values 
 are much larger than those obtained from the nonfrictional ratchet models 
 that were studied before. In the small parameter range of 
$\Gamma \sim 0.05, \, \phi \sim 0.2 , \, \alpha \sim 0.9 $ 
we find larger coherence with $Pe \, \sim \, 3.5 $. However, the 
transport coefficients are found to be very small. 
We have not considered here the mutual interplay between the thermal 
noise and the external noise which 
 may lead to an increase in coherence in appropriate parameter range. 
 Work along this line as well as on different frictional ratchet 
 models is under progress. 

 \end{document}